\documentclass[letter,traditabstract]{aa} 
\usepackage{graphicx}
\usepackage{txfonts}
\newcommand{\Msol}{$M_{\odot}$}
\newcommand{\Mjup}{$M_{\rm Jup}$}
\begin{document}
   \title{Infrared radial velocities of vB\,10}

   \author{M. R. Zapatero Osorio\inst{1,2}\fnmsep\thanks{e-mail: zapateroor@inta.es}       
          \and
          E. L. Mart\'\i n\inst{1,2,3}
          \and
          C. del Burgo\inst{4}
          \and
          R. Deshpande\inst{3}
          \and
          F. Rodler\inst{5}
          \and
          M. M. Montgomery\inst{3}
          }

   \institute{Centro de Astrobiolog\'\i a (CAB-CSIC), Ctra. Ajalvir km. 4, 
              E-28850 Torrej\'on de Ardoz, Madrid, Spain
          \and
              Formerly at Instituto de Astrof\'\i sica de Canarias, 
              E-38200 La Laguna, Tenerife, Spain
          \and
              University of Central Florida, Physics Department,  
              PO Box 162385, Orlando, FL32816, USA
          \and
              School of Cosmic Physics, Dublin Institute for Advanced Studies, Dublin 2, Ireland
          \and
              Instituto de Astrof\'\i sica de Canarias, 
              E-38200 La Laguna, Tenerife, Spain
   }

   \date{Received ; accepted }

  \abstract
  {We present radial velocities of the M8V-type, very low-mass star \object{vB\,10} that have been obtained at four different epochs of observations between 2001 and 2008. We use high-resolution ($R$\,$\sim$\,20,000) near-infrared ($J$-band) spectra taken with the {\sc nirspec} instrument on the Keck\,II telescope. Our data suggest that \object{vB\,10} shows radial velocity variability with an amplitude of $\sim$1 km\,s$^{-1}$, a result that is consistent with the recent finding of a massive planet companion around the star by Pravdo \& Shaklan (2009). More velocity measurements and a better sampling of the orbital phase are required to precisely constrain the orbital parameters and the individual masses of the pair.
  }


   \maketitle
%

\section{Introduction}
Over 300 planets are known orbiting main-sequence and giant stars of the solar vicinity (e.g., Udry \& Santos \cite{udry07}). Most  of them have been found or confirmed via the radial velocity technique (Mayor \& Queloz \cite{mayor95}; Marcy \cite{marcy98}). Recent  results suggest that less massive stars (M spectral types) tend to harbor less massive planets (e.g., Neptune-type and smaller). However, the report  by Pravdo \& Shaklan (\cite{pravdo09}) indicates that small stars can also have super-planets with masses several times larger than that of Jupiter. These authors present  the discovery of an unseen companion (of likely planetary-mass) around the low-mass star \object{vB\,10}, an object that is widely used as an M8V spectral standard star in the literature (Kirkpatrick et al. \cite{kirk95}; Mart\'\i n et al. \cite{martin96}). Actually, with a mass close to the frontier between stars and brown dwarfs, \object{vB\,10} is the least-massive star with a known planet. Wider planets have been found by direct imaging techniques around young brown dwarfs (Chauvin et al. \cite{chauvin04}; B\'ejar et al. \cite{bejar08}). \object{vB\,10} is also one of the closest low-mass systems ($d$\,=\,5.9 pc) to the Sun. Pravdo \& Shaklan's (\cite{pravdo09}) work on \object{vB\,10} is based on long-term astrometric observations, from which the authors derive the total mass of the pair to be 0.0841~\Msol~and estimate an orbital period of 0.744 yr. These authors discuss that the individual masses are likely 0.0779~\Msol~for the star and 6.4~\Mjup~for the planet; therefore, an amplitude of 1 km\,s$^{-1}$ or larger is expected for the radial velocity curve of the primary component (vB\,10\,A) of the system. Accurate radial velocities of \object{vB\,10} are needed to confirm or rule out the presence of the  massive planet. 

Here, we report on radial velocities obtained from near-infrared, high-resolution spectra. These velocities are measured to an accuracy below 1 km\,s$^{-1}$, which should be enough to study whether \object{vB\,10} shows variability and whether such velocity variability is consistent with the predictions of Pravdo \& Shaklan (\cite{pravdo09}) astrometric solution. \object{vB\,10} is brighter in the near-infrared wavelengths than in the optical ($R-J$\,$\sim$\,5.7 mag). In addition, this object is known to be active with flare events correlating with X-ray and UV emission (Berger et al. \cite{berger08}, and references therein). It has been discussed that radial velocity measurements of active cool dwarfs have the imprint of stellar activity if measured at optical wavelengths, while ``near-infrared'' values are less affected by a factor of at least 10 (Mart\'\i n et al. \cite{martin06}; Hu\'elamo et al. \cite{huelamo08}). We have recovered our past and recent near-infrared spectroscopic observations of \object{vB\,10} and we have analyzed them in a consistent way to derive radial velocities for different epochs that have been spread over the interval 2001---2008.


\section{Observations and data reduction}
As part of our program of radial velocity monitoring of late M-type stars (Deshpande et al. \cite{deshpande08}), we observed \object{vB\,10} on four different occasions between 2001 and 2008 using the Keck\,II telescope and the {\sc nirspec} instrument, which is a cross-dispersed, cryogenic echelle spectrometer employing a 1024\,$\times$\,1024 ALADDIN InSb array detector (McLean et al. \cite{mclean98}). The 2001 observations have been previously reported by us (Zapatero Osorio et al. \cite{osorio06,osorio07}), and the two recent epochs (2007 and 2008) are presented here for the first time. We provide the log of the observations in Table~\ref{log}, where the information related to the 2001 data is also included for completeness. In the echelle mode we selected the {\sc nirspec-3} ($J$-band) filter and an entrance slit width of 0.432\arcsec~(i.e., 3 pixels along the dispersion direction of the detector), except for the 2001 Jun observations for which we used an entrance slit width of 0.576\arcsec. The length of both slits was 12\arcsec. All observations were performed at an echelle angle of $\sim$63\degr. This instrumental setup provided a wavelength coverage from 1.148 up to 1.346\,$\mu$m split into 10 different echelle orders, a nominal dispersion ranging from 0.164 (blue) to 0.191 \AA\,pix$^{-1}$ (red wavelengths), and a final resolution element of 0.55--0.70 \AA~at 1.2485\,$\mu$m (roughly the central wavelength of the spectra), corresponding to a resolving power $R$\,$\sim$\,17,800 (2001 Jun) and 22,700 (remaining epochs). Individual exposure times were typically 100 or 120\,s for \object{vB\,10}. Weather conditions (seeing and atmospheric transparency) were fine during the observations, except for the 2008 epoch, which was hampered by cirrus and strong wind.

\begin{table}
\caption{{\sc nirspec} $J$-band observing log of vB\,10.\label{log}}
\centering                          
\begin{tabular}{lccccc}        
\hline\hline                 
Obs.~Date & UT & Slit name & Echelle & Exposure & Airmass \\
 & ($^h$ $^m$) &           &  (deg)  &   (s)    &         \\
\hline
2001 Jun 15 & 14:06 & 0.576$\times$12 & 63.00 & 2$\times$100 & 1.20       \\  
2001 Nov  2 & 04:43 & 0.432$\times$12 & 62.78 & 8$\times$120 & 1.16--1.22 \\  
2001 Nov  2 & 05:39 & 0.432$\times$12 & 63.29 & 8$\times$120 & 1.37--1.46 \\  
2007 Jun 25 & 13:22 & 0.432$\times$12 & 63.00 & 2$\times$120 & 1.18       \\  
2008 Jul 28 & 06:07 & 0.432$\times$12 & 63.00 & 2$\times$120 & 1.48       \\  
\hline
\end{tabular}
\end{table}

Raw data were reduced using the {\sc echelle} package within {\sc  iraf}\footnote{IRAF is distributed by National Optical Astronomy Observatory, which is operated by the Association of Universities for Research in Astronomy, Inc., under contract with the National Science Foundation.}. For consistency issues, the new observations and the 2001 data were reduced following the same procedure described in Zapatero Osorio et al. (\cite{osorio06}). Spectra were collected at two different positions along the entrance slit. Nodded images were subtracted to remove sky background and dark current. White light spectra obtained with the same instrumental configuration and for each observation of \object{vB\,10} were used for flat-fielding the data. Individual spectra were optimally extracted with the {\sc apall} task, and calibrated in wavelength using the internal arc lamp lines of Ar, Kr, and Xe, which were always acquired after observing \object{vB\,10} and before pointing the telescope to the next target. The air wavelengths of the arc lines were identified using the {\sc nist}\footnote{http://physics.nist.gov/PhysRefData/ASD/lines\_form.html} database, and we produced fits using a third-order Legendre polynomial along the dispersion axis and a second-order one perpendicular to it. The mean rms of the fits was 0.03 \AA, or 0.7 km\,s$^{-1}$. We note that this calibration method may produce systematic errors, or different zero-point shifts in velocity, that have to be removed before deriving precise radial velocities for \object{vB\,10}. Individual spectra were combined to produce one spectrum for each observing epoch and instrumental setup. In order to correct for atmospheric telluric absorptions, near-infrared featureless stars of spectral types A0--A2 (\object{HD\,181414}, \object{HD\,123233}, and \object{HD\,189920}) were observed close to the observations of our target, although they were not always acquired at the same airmass. Intrinsic lines to these hot stars, like strong hydrogen at 1.282 $\mu$m, were removed from the spectra before using them for division into the corresponding science data. Finally, we multiplied the science spectra of \object{vB\,10} by the blackbody spectrum for the temperature of 9480\,K, which is adequate for A0V type (Allen \cite{allen00}). All of our final spectra of \object{vB\,10} are characterized by a high signal-to-noise (s/n) ratio (s/n estimated at $\ge$50 for the red wavelenghts), except for the 2008 data, which have a factor of 3--5 lower quality.

\begin{figure}
\centering
\includegraphics[width=8.5cm]{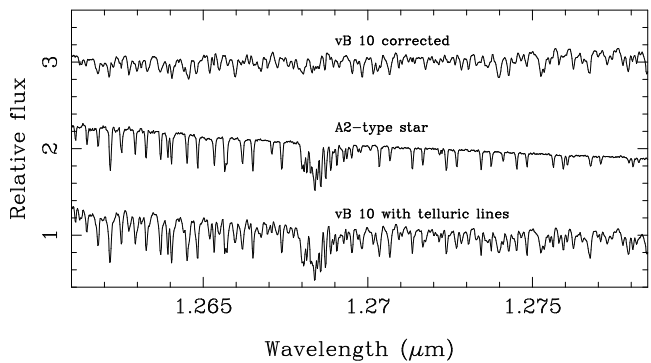}
\caption{{\sc nirspec} spectra (2007 Jun) of vB\,10 corresponding to the echelle order 60 (fully dominated by telluric line absorption). The top and bottom data correspond to vB\,10 after and before correction for the telluric contribution. The middle spectrum is the hot, featureless star \object{HD\,181414} taken on the same date. No velocity shift has been applied, and data are wavelength calibrated in the air system. The total spectral flux is normalized to unity and spectra are shifted vertically by 1. \label{spec}}
\end{figure}

\section{Radial velocities}
To investigate whether \object{vB\,10} shows radial velocity variability, we have measured the object's velocity displacement of all observing epochs relative to a fixed reference, which we have chosen to be the 2007 epoch spectrum because of its high quality. The {\sc nirspec} data were corrected for any wavelength shift with respect to the reference epoch. To do this, we have used the Earth's atmosphere telluric lines imprinted on the data; many of these lines are seen in sharp absorption and they profusely show in the echelle orders 66, 60, 58, and 57 (from blue to red wavelengths, for the echelle order numbering see Mclean et al. \cite{mclean07}). Figure~\ref{spec} depicts the spectra of \object{vB\,10} and one hot star corresponding to the echelle order 60. The telluric signal is markedly apparent in the spectra of \object{vB\,10} before division by the standard hot star, and it clearly dominates over the target signal. The radial velocity of the telluric lines is constant in all wavelengths down to a level of $\sim$$\pm$10 m\,s$^{-1}$ (e.g., Seifahrt \& K\"aufl \cite{seifahrt08}), which is $\ge$10 times smaller than the velocity precision we can achieve with {\sc nirspec}. Such a constancy is applicable over all epochs. These lines thus become excellent calibrators of the wavelength zero points and they can be employed to probe the intrinsic stability of the {\sc nirspec} spectrograph. 

We have cross-correlated the echelle orders heavily covered by telluric lines against the 2007 epoch using the task {\sc fxcor} within {\sc iraf}. At this step, we cross-correlate only spectra uncorrected for the telluric contribution. Because of the good quality of our data and the marked strength of the telluric features, the cross-correlation functions show clearly defined narrow peaks. The velocity displacements are derived from the center of these peaks. The last column of Table~\ref{data} provides the shifts and their associated uncertainties, which are mathematically computed from the errors of the mean shifts. These errors are typically around 200--300 m\,s$^{-1}$. Two epochs (2001 Jun and the first spectrum of 2001 Nov) appear to share the same wavelength zero point, within 1-$\sigma$ uncertainty, than the reference epoch. 

To measure the relative radial velocities of \object{vB\,10}, we have cross-correlated the telluric-free echelle orders 62 (central wavelength at 1.230 $\mu$m), 61 (1.250 $\mu$m), and 59 (1.2925 $\mu$m) against the 2007 epoch. The cross-correlation functions show a neat peak since \object{vB\,10} itself acts as the template. The derivations for the second spectrum of 2001 Nov and 2008 Jun have been compensated for the different wavelength zero points. Our results, obtained from the average of the various echelle orders, are given as a function of the Modified Julian Date in the third column of Table~\ref{data}. These values have been properly corrected for the diurnal (Earth's rotation), lunar (motion of the Earth's center about the Earth-Moon barycenter), and annual (motion of the Earth-Moon barycenter about the Sun) velocities. As a test of consistency, we have also cross-correlated the telluric-corrected spectra of \object{vB\,10}, including the echelle orders that were heavily ``obscured'' by the telluric lines, deriving very similar values within 1-$\sigma$ errors. The uncertainties associated to the relative radial velocities are computed from the additive combination of the wavelength zero point shifts error bars (last column of Table~\ref{data}) and the errors of the mean relative velocities. 

\begin{table}
\begin{minipage}[t]{\columnwidth}
\caption{{\sc nirspec} radial velocity measurements of vB\,10.\label{data}}
\centering
\renewcommand{\footnoterule}{}  
\begin{tabular}{lccc}
\hline \hline
MJD & $v_r$ & Relative $v_r$ & $v_{\rm shift}$\footnote{Relative offset in velocity zero point.} \\
   & (km\,s$^{-1}$) & (km\,s$^{-1}$) & (km\,s$^{-1}$) \\ 
\hline
52075.58951 &  35.50 $\pm$ 0.60 &  $+$1.13 $\pm$ 0.30  &  $-$0.08 $\pm$ 0.23 \\  
52215.20560 &  34.45 $\pm$ 0.60 &  $+$0.08 $\pm$ 0.30  &  $-$0.17 $\pm$ 0.22 \\  
52215.24225 &  34.40 $\pm$ 0.60 &  $+$0.03 $\pm$ 0.30  &  $+$0.51 $\pm$ 0.23 \\
54276.55865\footnote{Reference epoch (see text).} &  34.37 $\pm$ 0.30 & -- &   --       \\   
54675.25669 &  34.00 $\pm$ 0.83 &  $-$0.37 $\pm$ 0.53  &  $+$1.07$\pm$  0.24 \\
\hline
\end{tabular}
\end{minipage}
\end{table}

The two spectra of \object{vB\,10} taken in 2001 Nov, separated by $\sim$50 min, indicate that the star had a ``constant'' velocity (within $\pm$300 m\,s$^{-1}$) in that short-time interval. However, the first epoch measurement shows a non-zero relative radial velocity, suggesting some variability with an amplitude of around 1 km\,s$^{-1}$. The significance of such variability is estimated at the 3.2-$\sigma$ level when compared to the mean of all velocity measurements. 

To derive the absolute heliocentric velocities of \object{vB\,10} we need to correct the reference epoch for any possible zero point offset in its corresponding wavelength calibration solution. We have used the Earth transmission spectrum above Mauna Kea generated by the ATRAN modelling software (Lord \cite{lord92}) in the wavelength interval of our observations as a template in the cross-correlation of the 2007 spectrum of \object{vB\,10} (only echelle orders containing strong telluric absorption lines). A velocity shift of $-$0.88 km\,s$^{-1}$ is found with a dispersion of $\pm$0.27 km\,s$^{-1}$ (error of the mean\,=\,0.14 km\,s$^{-1}$). The heliocentric radial velocity is then computed by measuring the centroids of 11 atomic lines due to K\,{\sc i}, Fe\,{\sc i}, Ti\,{\sc i}, and Mn\,{\sc i} (McLean et al. \cite{mclean07}), and by correcting for the diurnal, lunar and annual velocities, obtaining $v_r$\,=\,34.37\,$\pm$\,0.30 km\,s$^{-1}$ for the reference 2007 epoch. The uncertainty is derived as the mean error of the observed velocity (atomic line centroids) plus the mean error of the velocity zero point shift. The final heliocentric velocities for all epochs are given in the second column of Table~\ref{data}. 

Our heliocentric velocity measurements confirm that the ``mean'' radial velocity of \object{vB\,10} is indeed near 35 km\,s$^{-1}$ as it has been frequently claimed in the literature (e.g., Tinney \& Reid \cite{tinney98}; Mart\'\i n \cite{martin99}; Basri \& Reiners \cite{basri06}; Zapatero Osorio et al. \cite{osorio07}). We note that the heliocentric radial velocities of \object{vB\,10} given in Table~1 by Zapatero Osorio et al. (\cite{osorio07}), which correspond to the first two lines of Table~\ref{data}, were not corrected for any wavelength zero point offset. Pravdo \& Shaklan (\cite{pravdo09}) also referred to these measurements out of a total of six published velocities obtained in the year interval 1992--2001. Interestingly, the velocity difference between the first two epochs 2001 Jun and 2001 Nov was already apparent in Zapatero Osorio et al. (\cite{osorio07}), but the associated error bars were too large for a firm assesment. Presently, we have reduced the uncertainties by applying careful zero point corrections in the wavelength solutions and by using a large number of spectroscopic features in the cross-correlation technique in contrast to the few lines (only K\,{\sc i}) used for \object{vB\,10} in Zapatero Osorio et al. (\cite{osorio07}). Unfortunately, the remaining four velocities compiled by Pravdo \& Shaklan (\cite{pravdo09}, see their Table~4), which include those by Basri \& Reiners (\cite{basri06}), who claimed that \object{vB\,10} is a star with probably constant velocity based on a velocity dispersion of 0.3\,$\pm$\,1.4 km\,s$^{-1}$, are not precise enough to further constrain the nature of the M8V-type star.

An alternative method to obtain the absolute heliocentric velocity of \object{vB\,10} is via the comparison of the observed spectra to theoretical spectra. Following the procedure described in del Burgo et al. (\cite{burgo09}), we have found that \object{vB\,10} is well matched by the $T_{\rm eff}$\,=\,2700\,$\pm$\,250 K, log\,$g$\,=\,4.8\,$\pm$\,0.5 [cm\,s$^{-2}$], solar metallicity models (further details will be provided in del Burgo et al., in preparation). The wavelength cross-correlation of the 2007 spectrum of \object{vB\,10} against the synthetic data yields an heliocentric velocity of 33.17\,$\pm$\,0.17 km\,s$^{-1}$, where the uncertainty stands for the standard deviation of the radial velocities obtained from the various synthetic models within the 1-$\sigma$ error range of the atmospheric parameters. After applying the velocity zero point offset, the final value is 34.05\,$\pm$\,0.30 km\,s$^{-1}$. This determination compares with the one of Table~\ref{data}, and both measurements are consistent within 1-$\sigma$ uncertainty, indicating that our procedure for measuring velocities is reliable.

\begin{figure}
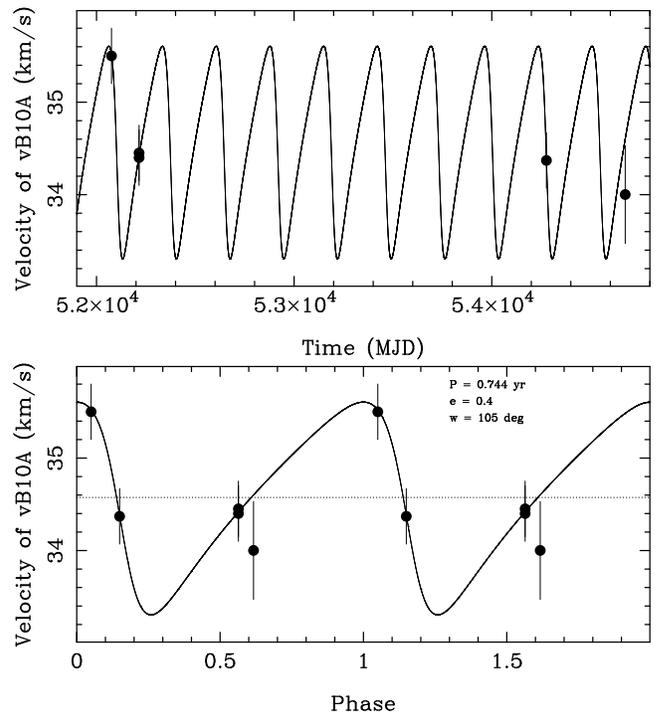

\centering
\includegraphics[width=8.5cm]{rv_v2.ps}
\includegraphics[width=8.5cm]{phase.3.ps}
\caption{Our radial velocity measurements (filled dots) are plotted against the Modified Julian Date of the observations (top panel), and the orbital phase (bottom panel). Overplotted onto the data is the ``most likely'' astrometric Keplerian solution by Pravdo \& Shaklan (\cite{pravdo09}), for which we have assumed an orbital eccentricity of 0.4 and an argument of periapsis of 105\degr. These spectroscopic observations cannot constrain the orbital solution precisely, but they are consistent with the presence of a small (planetary-mass) body aroud vB\,10. The horizontal dotted line denotes the ``systemic'' velocity of the pair. Note that two periods are depicted in the bottom diagram. \label{rv}}
\end{figure}

\section{Discussion and final remarks}
The {\sc nirspec} velocities of \object{vB\,10} (Table~\ref{data}) provide a hint for the presence of variability with a possible amplitude of $\sim$1 km\,s$^{-1}$. However, we caution that this is supported by one epoch measurement, which deviates by 3.2-$\sigma$ from the remaining three epochs. This appears insufficient to unambiguously confirm the planet around \object{vB\,10} since there is no evidence for a periodic signal in the spectroscopic data. It would be desirable to have more velocity measurements for a definitive analysis on the velocity variability, amplitude, and periodicity. Our results do not suggest that \object{vB\,10} contains a stellar-mass or a heavy brown-dwarf-mass companion in a short, not very eccentric orbit.

The origin of the $\sim$1 km\,s$^{-1}$ amplitude may be due to stellar activity, uncontrolled systematic errors in our measurements, or the presence of a massive planet. In the case of the active, young, M9-type brown dwarf \object{LP\,944$-$20}, near-infrared velocity variations are below 360 m\,s$^{-1}$ (Mart\'\i n et al. \cite{martin06}). Yet, we do not know the impact of flares on the determination of velocities. \object{vB\,10} shows X-ray flares at a rather low rate, which we have determined to be 2\%~of the time for the strong events (log\,$L_X/L_{\rm bol}$ $\ge -3$) and 3\%~for the faint ones (log\,$L_X/L_{\rm bol}$ $\sim -4$, Linsky et al. \cite{linsky95}; Fleming et al. \cite{fleming93}, \cite{fleming00}, \cite{fleming03}; Berger et al. \cite{berger08}), and similar flare duty cycle frequencies ($\sim$0.7\%~and 6\%) were derived for \object{LP\,944$-$20} by Mart\'\i n \& Bouy (\cite{martin02}). Although the probability of observing \object{vB\,10} while flaring is small ($\sim$5\%), we cannot discard that the measured velocity change is influenced by a strong atmospheric activity during the first observing epoch.

We now focus on whether our velocities agree with or refute the predicted velocity curve of vB\,10\,A obtained from the ``most likely'' astrometric solution and mass estimates of Pravdo \& Shaklan (\cite{pravdo09}). Summarizing, these authors discuss that \object{vB\,10} comprises a primary object near the substellar borderline (mass of 0.0779~\Msol) and a massive planet (6.4~\Mjup), both of which orbit around the center of mass of the system with a periodicity of 0.744 yr. The orbital separation between the two pair members is astrometrically measured at 0.36 AU. Additional orbital parameters are provided in Table~5 of Pravdo \& Shaklan (\cite{pravdo09}). We note that the orbital eccentricity ($e$), epoch, and argument of periastron ($w$) are not constrained by the astrometric solution. Our spectroscopic data are sparse and insufficient to provide additional strong restrictions to the orbital solution. However, from the visual inspection of several tens of computed velocity curves produced by varying the orbital epoch, $e$, and $w$, we infer that $e$\,$\le$\,0.8 and $w$\,$\sim$\,10\degr--160\degr~for epochs in the range 52070--52150 (MJD) may provide a reasonable match to the observations.

The top panel of Fig.~\ref{rv} shows the {\sc nirspec} velocities as a function of the observing time and an example of a predicted velocity curve computed for $e$\,=\,0.4, $w$\,=\,105\degr, and epoch\,=\,52106 (MJD), which minimizes the expression $\sum[|v_{\rm computed} - v_r|$/err($v_r$)]. This eccentricity is very close to the one derived from the solution of the combined astrometric and {\sc nirspec observations performed by Pravdo \& Shaklan (priv. communication).} The bottom panel depicts the velocity curve folded in phase with a periodicity of 0.744 yr. As regarding the systemic velocity of \object{vB\,10}, we have adopted the mean value of all {\sc nirspec} velocities: 34.57 km\,s$^{-1}$, with an uncertainty of 0.30 km\,s$^{-1}$. A relatively high eccentricity of $e$\,$\le$\,0.8 is found in many giant planets orbiting solar-type stars (e.g., Udry \& Santos \cite{udry07}; Butler et al. \cite{butler06}) as well as in the brown dwarf--low-mass star pair GJ\,569Bab (Lane et al. \cite{lane01}); therefore, it appears to be common in Nature and indicates that low-mass (planetary) companions can be found with a rich variety of orbital shapes around stars independently of the mass ratio of the system. As illustrated in Fig.~\ref{rv}, the agreement between the velocity curve prediction and the observations is remarkably within 1-$\sigma$ the uncertainties giving some credit to the mass estimates and the astrometric solution obtained by Pravdo \& Shaklan (\cite{pravdo09}). 

More radial velocity measurements with a better time sampling of the orbital period in addition to the astrometric data are highly required for a detailed knowledge of the orbital parameters of the pair vB\,10\,Ab. We have shown that {\sc nirspec} can provide radial velocities with error bars of $\ge$\,0.2 km\,s$^{-1}$, but significantly more accurate determinations (with uncertainties a factor of 10 smaller) or many more {\sc nirspec} measurements would be necessary for the precise characterization of the individual masses of this planetary system.

The discovery of a likely massive planetary companion to a very low-mass star like \object{vB\,10} (planet-to-star mass ratio of $\sim$0.08) inside the region where circum(sub)stellar disks are present at very young ages adds a new constraint to models of planet formation. This together with the previous findings of wide planetary-mass companions to brown dwarfs (e.g., Chauvin et al. \cite{chauvin04}; B\'ejar et al. \cite{bejar08}) indicates that gravitational instabilities may play a role on the formation of super-planets and/or that these systems may also form like binary stars.

\begin{acknowledgements}
We are grateful to R. Tata, H. Bouy, and P.-B. Ngoc, and to N. Dello Russo for helping to obtain the 2007 and 2008 {\sc nirspec} spectra, respectively. We thank S. Witte, P. Hauschildt, and Ch. Helling for providing computer-ready synthetic spectra. We thank the Keck observing assistants and the staff in Waimea for their kind support. Some data were also obtained at the W. M. Keck Observatory, which is operated as a scientific partnership between the California Institute of Technology, the University of California, and NASA. The Observatory was made possible by the generous financial support of the W. M. Keck Foundation. The authors extend special thanks to those of Hawaiian ancestry on whose sacred mountain we are privileged to be guests. This work is partly financed by the Spanish Ministry of Science through the project AYA2007--67458 and by a Keck data grant provided by NASA through the Michelson Science Center. 
\end{acknowledgements}


\begin{thebibliography}{}
\bibitem[2000]{allen00}{Allen, W. B. 2000, in Allen's Astrophysical 
         Quantities, ed. A. N. Cox (4th ed; New York: Springer), 151}
\bibitem[2008]{bejar08}{B\'ejar, V. J. S., Zapatero Osorio, M. R., 
         P\'erez-Garrido, A. et al.  2008, \apj, 673, L185}
\bibitem[2008]{berger08}{Berger, E., Basri, G., Gizis, J. E. 
         et al. 2008, \apj, 676, 1307}
\bibitem[2006]{butler06}{Butler, R. P., Wright, J. T., Marcy, G. W. 
         et al. 2006, \apj, 646, 505}
\bibitem[2006]{basri06}{Basri, G., \& Reiners, A. 2006, \apj, 132, 663}
\bibitem[2009]{burgo09}{del Burgo, C., Mart\'\i n, E. L., Zapatero Osorio, 
         M. R., \& Hauschildt, P. H.  2009, \aap, 501, 1059}
\bibitem[2004]{chauvin04}{Chauvin, G., Lagrange, A.- M., Dumas, C., 
         et al. 2004, \aap, 425, L29}
\bibitem[2008]{deshpande08}{Deshpande, R., Mart\'\i n, E. L., Montgomery, 
         M. M., et al. 2008, AAS meeting abstract, 211, 159.03}
\bibitem[1993]{fleming93}{Fleming, T. A., Giampapa, M. S.,   
         Schmitt, J. H. M. M., \& Bookbinder, J. A. 1993, \apj, 410, 387}
\bibitem[2000]{fleming00}{Fleming, T. A., Giampapa, M. S., \&
         Schmitt, J. H. M. M. 2000, \apj, 533, 372}
\bibitem[2003]{fleming03}{Fleming, T. A., Giampapa, M. S., \&
         Garza, D. 2003, \apj, 594, 982}
\bibitem[1995]{linsky95}{Linsky, J. L., Wood, B. E., Brown, A., 
         Giampapa, M. S., \& Ambruster, C. 1995, \apj, 455, 670}
\bibitem[1992]{lord92}{Lord, S. D. 1992, NASA Technical 
         Memor. 103957}
\bibitem[2008]{huelamo08}{Hu\'elamo, N., Figueira, P., Bonfils, X. 
         et al. 2008,  \aap, 489, L9}
\bibitem[1995]{kirk95}{Kirkpatrick, J. D., Henry, T. J., \& Simons, 
         D. A. 1995, \aj, 109, 797}
\bibitem[2001]{lane01}{Lane, B. F., Zapatero Osorio, M. R., Britton, 
         M. C., Mart\'\i n, E. L., \& Kulkarni, S. R. 2001, \apj, 560, 
         390}
\bibitem[1998]{marcy98}{Marcy, G. W. 1998, \nat, 391, 127}
\bibitem[1995]{mayor95}{Mayor, M., \& Queloz, D. 1995, 
         \nat, 378, 355}
\bibitem[2006]{martin06}{Mart\'\i n, E. L., Guenther, 
         E., Zapatero Osorio, M. R.,  Bouy, H., \& Wainscoat, R. 
         2006, \apj, 644, L75}
\bibitem[1996]{martin96}{Mart\'\i n, E. L., Rebolo, 
         R., \& Zapatero Osorio, M. R. 1996, \apj, 469, 706}         
\bibitem[1999]{martin99}{Mart\'\i n, E. L. 1999, \mnras, 302, 59}
\bibitem[2002]{martin02}{Mart\'\i n, E. L., \& Bouy, H. 2002, 
         \na, 7, 595}
\bibitem[1998]{mclean98}{McLean, I. S., Becklin, E. E., Bendiksen O., 
         et al. Proc. SPIE, 3354, 566}
\bibitem[2007]{mclean07}{McLean, I. S., Prato, L., McGovern, M. R., 
         et al. 2007, \apj, 658, 1217}
\bibitem[2009]{pravdo09}{Pravdo, S. H., \& Shaklan, S. B. 2009, \apj, 
         700, 623}
\bibitem[2008]{seifahrt08}{Seifahrt, A., \& K\"aufl, H. U. 2008, \aap, 
         491, 929}
\bibitem[1998]{tinney98}{Tinney, C. G., \& Reid, I. N. 1998, \mnras, 
         301, 1031}
\bibitem[2007]{udry07}{Udry, S., \& Santos, N. C. 2007, \araa, 45, 397}
\bibitem[2006]{osorio06}{Zapatero Osorio, M. R., Mart\'\i n, E. L., 
         Bouy, H., et al. 2006, \apj, 647, 1405}
\bibitem[2007]{osorio07}{Zapatero Osorio, M. R., Mart\'\i n, E. L., 
         B\'ejar, V. J. S., et al. 2007, \apj, 666, 1205}

\end{thebibliography}
\end{document}